\documentclass[twocolumn]{aastex631}

\usepackage{hyperref}
\usepackage{verbatim}
\submitjournal{ApJL}
\usepackage{amsmath}
\usepackage{amsfonts}
\shorttitle{PSR J0952$-$0607's Tightened Mass}
\shortauthors{Romani et al.}
\defcitealias{romani2011orbit}{RS11} 	
\defcitealias{Kandel_2020}{KR20}
\defcitealias{2017ApJ...845...42S}{SR17}

\begin{document}

%\title{PSR J0952-0607: An Improved Light Curve and Decreased $\delta M$}
\title{PSR J0952-0607: Tightening a Record-High Neutron Star Mass}

\correspondingauthor{R. W. Romani}
\email{rwr@astro.stanford.edu}

\author[0000-0001-6711-3286]{Roger W. Romani}
\affil{Department  of  Physics,  Stanford  University,  Stanford,  CA 94305, USA}
\author[0000-0003-1285-8170]{Maya Beleznay}
\affil{Department  of  Physics,  Stanford  University,  Stanford,  CA 94305, USA}
\author[0000-0003-3460-0103]{Alexei V. Filippenko}
\affil{Department of Astronomy, University of California, Berkeley, CA 94720-3411, USA}
%\affil{Miller Institute for Basic Research in Science, University of California, Berkeley, CA 94720, USA}
\author{Thomas G. Brink}
\affil{Department of Astronomy, University of California, Berkeley, CA 94720-3411, USA}
\author{WeiKang Zheng}
\affil{Department of Astronomy, University of California, Berkeley, CA 94720-3411, USA}

\begin{abstract}
We report on new orbit-minimum photometry and revised radial-velocity fitting that provide an improved measurement of the mass of the neutron star (NS) in pulsar PSR~J0952$-$0607 at $M_{\rm NS} = 2.35\pm 0.11\, M_\sun$. With its fast spin and unusually low magnetic field, this NS has evidently experienced unusual evolution, likely connected with its high mass, which is now $2.5\sigma$ above that of the heaviest pulsar with a white dwarf companion, as measured by Shapiro delay techniques. By tightening the mass measurement, we also raise the maximum (commonly called Tolman-Oppenheimer-Volkoff) NS mass to $M_{\rm TOV} > 2.27\,M_\odot$\,$(2.12\,M_\odot)$ at $1\sigma$\,$(3\sigma)$ confidence, which improves bounds on the dense-matter equation of state. While the statistical error decreases and systematic issues should be modest, uncertainties remain; we comment briefly on these factors and prospects for further improvement.
\end{abstract}


\section{Introduction} \label{sec:intro}

Pulsar PSR~J0952$-$0607 (hereafter J0952) is the fastest spinning Galactic disk pulsar with $P_s=1.41$\,ms and, with a \citet{1970SvA....13..562S}-corrected ${\dot P_{\rm obs}} = 4.6 \times 10^{-21}\,{\rm s\,s^{-1}}$, has among the lowest pulsar surface dipole fields at $8.2\times 10^7$\,G \citep{2017ApJ...846L..20B}. As a black-widow pulsar, it heats the face of its low-mass ($0.03\,M_\odot$) companion. \citet[][hereafter R22]{2022ApJ...934L..17R} reported on model fits to Keck imaging and spectroscopy of this companion which gave a large neutron star (NS) mass of $M_{\rm NS} = 2.35\pm 0.17\, M_\sun$. 

While ``spider'' (black-widow/redback) mass measurements depend on optical modeling and are thus less ``clean'' than pure dynamics estimates from pulsar timing using general relativistic effects, such precision timing is typically  achieved only for pulsars with compact [white dwarf (WD) or NS] companions. However, the spiders (with longer, slower mass-transfer phases) may approach the Tolman-Oppenheimer-Volkoff limit $M_{\rm TOV}$, the maximum mass allowed by the dense-matter equation of state \citep[EOS;][]{alsing2018evidence}. Thus, precision measurements of these objects can be particularly important. J0952 plays a special role in this quest. Its modest heating flux and low companion fill factor make this object relatively immune from model systematics, at the cost of a low flux that makes the observations challenging. The modest heating ensures that the source is nearly flare-free and that the systematic uncertainties due to the hot spots, winds, and other factors present in other spider binaries with stronger companion heating are not present. A simple direct-heating model provides the best fit, minimizing model systematics. Since the EoS model constraining the power of this record-high mass depends on its lower bound, it is desirable to tighten the measurement. We report here on new Keck photometry which, along with spectrum reanalysis, allow a significant improvement.

\section{2025 Keck Observations}

The R22 statistical mass uncertainty, $\delta M_{\rm NS}=0.17\,M_\odot$, stems from imprecision in the companion center-of-mass radial velocity $K_{\rm CoM}$ and in the orbital inclination $i$. Inclination is constrained by the orbital light curve, especially via the flux and color at minimum brightness. Since existing (Semesters 2018A, 2019A) Keck-I 10\,m LRIS \citep{oke1995keck} photometry suffered from mediocre seeing, we have sought improved LRIS $g/I$ imaging under better conditions. While several Keck half-nights were scheduled (chosen so that orbital minimum occurred with low airmass and lunation), no periods of the subarcsec seeing required for this very challenging observation occurred. 
% In the event 
Thus, we adopted a ``target-of-opportunity'' strategy and computed the critical phase windows for all nights that A.V.F.'s team had available for various projects -- and then switched to J0952 when conditions became sufficient during the key phase. 
Through the flexibility of our colleagues, 
we were able to use LRIS to obtain $12\times 650$\,s~$g$-band + $24\times 300$\,s~$I$-band exposures on February 27, 2025 (MJD 60733.437-3.540) and $7\times 650$\,s~$g$ + $14\times 300$\,s~$I$ on April 27, 2025 (MJD 60792.304-2.357). The 0.75--0.85$''$ seeing during these observations allowed good photometry across orbital minimum brightness.

\begin{figure}[h!]
\centering
    \vspace*{-18mm}\hspace*{-10mm}\includegraphics[scale=0.48]{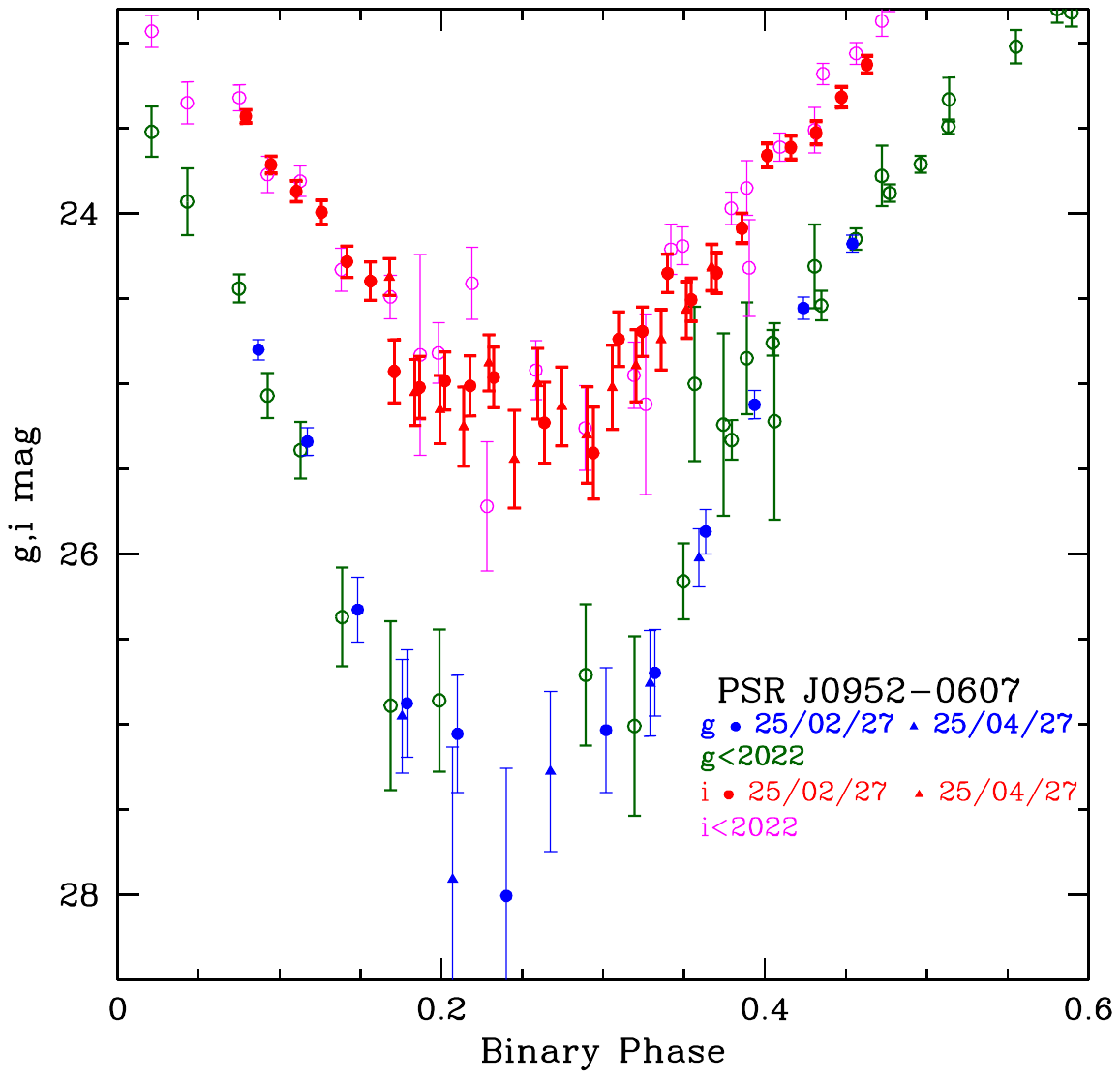}
    \vspace*{-30mm}
\caption{Keck LRIS $g/i$ photometry near orbital minimum brightness.}
\label{fig:LCmin}
\end{figure}

The simultaneous $g/I$ images were calibrated, and forced point-spread-function (PSF) relative photometry was performed at the positions of the pulsar and a grid of comparison stars with catalog PS2 magnitudes. The resulting pulsar photometry was then converted to the SDSS system \citep{finkbeiner2016hypercalibration}. One $g$ frame suffered a cosmic ray at the pulsar position and a second near minimum brightness did not give a reliable pulsar detection; all other frames produced acceptable measurements. These magnitudes are plotted in Fig.~\ref{fig:LCmin}, along with the $g/i$ photometry of R22. The agreement between the two nights (and with the archival photometry) is excellent, with the new data providing substantially more phase points with lower scatter, indicating that we have successfully measured the orbital minimum in these two filters near quiescence. With a minimum at $g>27$\,mag, the good seeing was clearly essential to success. Note that, while flaring activity is evidently very low in J0952, a few points rise above the general trend, especially in the archival data. These may be excised for the best fits to a quiescent light curve.

\section{Photometric/Radial-Velocity Fitting}

To fit the light curve we use the {\tt ICARUS} \citep{breton2012koi} code with added physics  \citep{Kandel_2020,2021ApJ...908L..46R}. The fit parameters are the orbital inclination $i$, the companion Roche lobe fill factor $f_1$, the heating luminosity $L_H$, the companion's night-side temperature $T_N$, and the system distance $d$. There can also be small offsets for ``bandpass calibrations," owing to possible imperfections in the photometric zero points. The free parameters are fit with Bayesian parameter estimation from Multinest sampling \citep{feroz2009multinest} using its {\tt Python} implementation {\tt pymultinest} \citep{buchner2016pymultinest}.  In addition to the photometry, the fits use the pulsar kinematic parameters of \citet{2019ApJ...883...42N}. 

We add the new $g$ and $i$ data to the $ugriz$ photometry of the R22 ``trimmed'' data, which excises a few clear photometric outliers. The fits proceed as in the previous analysis, giving fully consistent parameters, with reduced errors, including for $i$. While there are no clear flares in the new data, there are a few small positive fluctuations near minimum brightness, when the companion is faintest, leading to a reduced $\chi^2$ slightly larger than 1. These can increase the inferred minimum flux, biasing $i$ low and increasing the inferred NS mass. We found seven points with $>2.5\sigma$ when four were expected, so we optionally trimmed these seven outliers and refit. This results in slightly higher $i$, smaller uncertainties, and reduced $\chi^2<1$. The difference is small, but we adopt this higher $i$ solution as our fiducial fit, to be conservative. Parameters and uncertainties for the two fits (Table \ref{table:fit}) are well within $1\sigma$ of each other and of the values reported by R22. 

\begin{deluxetable}{lcc}[b!!]
\tabletypesize{\footnotesize}
\tablewidth{0pt}
\tablecaption{Light-Curve/RV-Fit Results for J0952$^a$\label{tab:lc_fit}}
\tablehead{
\colhead{Parameters} & \colhead{Trimmed} & \colhead{All}}
\startdata
$i\, (\mathrm{ deg})$&$60.2^{+1.2}_{-1.1}$&$60.1\pm1.2$\\
$f_1$ & $0.79\pm 0.01$ & $0.79\pm 0.01$ \\
$L_{\mathrm{H}}\,/10^{34}\,(\mathrm{erg/s})$ & $3.76^{+0.42}_{-0.40}$  &$3.79^{+0.42}_{-0.41}$\\
$T_{\rm N}$\,(K) & $3039^{+52}_{-55}$ & $3049\pm54$ \\
$d_{\rm kpc}$ & $6.26^{+0.40}_{-0.47}$ & $6.29^{+0.38}_{-0.46}$ \\
 $\chi^2/{\rm DoF}$ & 309/(346-10)[0.92] &359/(353-10)[1.05] \\
 \hline
 \hline
 $K_{\rm CoM}$\,(km/s) & $377.6\pm 3.8$ & $387.9\pm 6.6$ \\
 $M_{\rm NS}\,(M_\odot)$ & $2.347\pm 0.107$ & $2.549\pm 0.157$  \\
 $M_{\rm C}\,(M_\odot)$ &  $0.032\pm 0.001$ & $0.034\pm 0.002$ \\
 $\chi^2/{\rm DoF}$ &71/(48-2)[1.5]& 314/(54-2)[6.0]\\
\enddata
 \tablenotetext{}{$^a$ $A_V=0.16$\,mag; also fitted $\Delta u$, $\Delta g$, $\Delta r$, $\Delta i$, $\Delta z$.}
\label{table:fit} 
\end{deluxetable}

We next treat the data from six 2018--2022 Keck LRIS spectral campaigns, using the calibrated, extracted, and archived spectra from R22. As in that paper, we monitor drifts in the wavelength calibration of the extracted spectra by measuring the radial velocity (RV) of a comparison star included in the long slit. For the first two epochs in Semesters 2018B and 2019A a bright M3V star was used, while for the later epochs a G1V comparison was used, which had colors much closer to those of the pulsar companion at maximum brightness. These RVs and those of the pulsar were measured by cross-correlation using the IRAF {\tt rvsao} package \citep{1998PASP..110..934K} with standard-star templates. The velocity relative to the comparison stars was fit to determine the binary RV amplitude $K$. Since the companion center of light (CoL) lies toward the heated face from the center of mass (CoM), and since this offset depends on the disposition of the lines dominating the RV measurement, one requires a correction $K_{\rm cor}$ to scale the observed $K_{\rm CoL}$ to the desired $K_{\rm CoM}$. This correction depends on the companion heating pattern and the line features (or equivalently template spectral class) used. R22 used a G1V template for the companion; from the metal-line equivalent-width temperature dependence we estimated $K_{\rm cor}=1.05$. 

However, we can use the ICARUS photometric solution to compute companion model spectra, by integrating BT-Settl atmosphere emergent spectra \citep[][and references therein]{2013MSAIS..24..128A} over the visible face of the companion star for the phase of each observed spectrum, including all heating, limb darkening, and gravity darkening effects. This produces spectra consistent with the ICARUS model and correctly follows the varying line shifts (and line-shape variations) for each element of the companion surface, incorporating the offsets from the companion's CoM, and modeling nonsinusoidal components of the observed velocities. This automatically and self-consistently incorporates the heating pattern, obviating the need to estimate $K_{\rm cor}$. In principle, it also provides an ideal multitemperature template to match the companion spectrum at each phase, and thus should result in improved cross-correlations and better RVs. 

\begin{figure}[t!]
\centering
\vspace*{-15mm}\hspace*{-12mm}   
\includegraphics[scale=0.5]{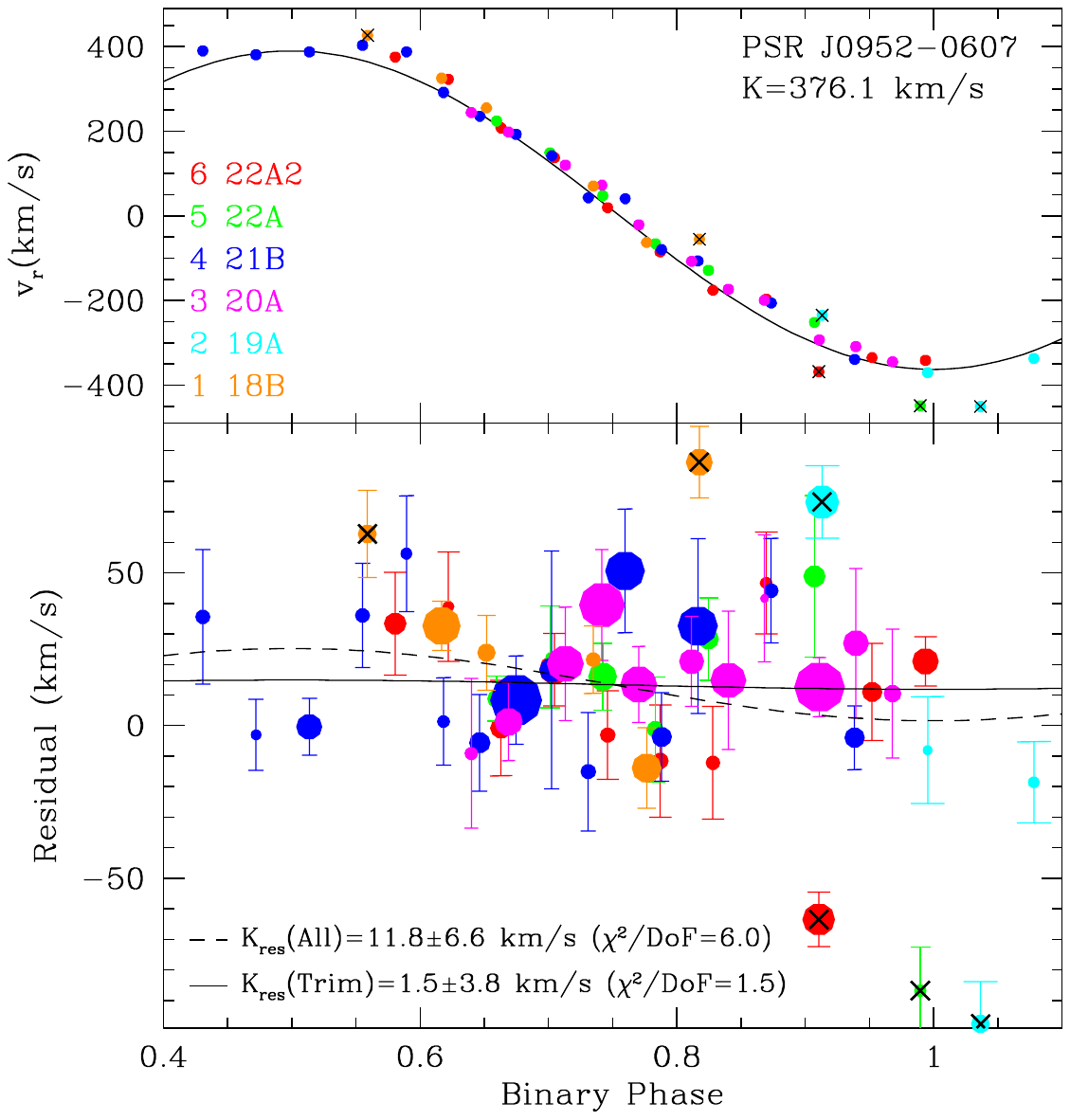}
    \vspace*{-30mm}
\caption{J0952 RV fits to Keck/LRIS spectra. The upper panel shows the full RV, while the lower panel shows the residual to the R22 RV model. The point size is scaled to the correlation coefficient $R$; a number of clear outliers marked with a black ``x'' are excised in a trimmed fit.}
\label{fig:RVfits}
\end{figure}

We therefore used the $K_{\rm CoM} = 376.1$\,km\,s$^{-1}$ RV measurement of R22 to compute model spectra for each phase and cross-correlated the observed spectra with these models to fit for residual RVs. A sinusoid fit to velocities measured for the computed models with a G1V template in fact finds $K_{\rm cor}=1.070$ with a nonsinusoidal component of amplitude $\sim 5$\,km\,s$^{-1}$, so fits to model spectra should be appreciably more reliable.

If the R22 value were perfect, we would expect to fit a $\Gamma$ offset for the poorly known absolute RV relative to the average of the comparison stars, but a zero RV amplitude. Following R22, we cross-correlated over the 3800--10,000\,\AA\ range, resampled to 8192 wavelengths, and filtered in Fourier space, typically with a lower cosine-bell filter cutoff from modes 10--100 and the upper cutoff running from modes 1800--3200. We expect a correlation peak near zero velocity, after correction for the calibration offset as monitored by the comparison star. As in our previous analysis, the strongest peaks usually corresponded to this low-residual solution. However, in a number of cases the expected correlation peak appeared as a secondary peak or shoulder to a separate cross-correlation maximum. The true peak could generally be isolated by an adjustment of the Fast Fourier Transform (FFT) taper, but for a few spectra, especially in the early 2018B and 2019A epochs, the spurious peak was always dominant. The fit velocity residuals are shown in Fig.~\ref{fig:RVfits}, with an upper panel displaying the full RVs for scale. We also show weighted least-squares fits to these residuals. Fitting all data returns a significantly increased RV amplitude, but the scatter is large with reduced $\chi^2/{\rm DoF}= 6$. However, most of this increase comes from evident outliers; if we trim the half dozen worst points, we get a better (albeit imperfect) scatter. The RV is smaller, consistent with that of R22, so we adopt it for our conservative mass solution.

\begin{figure}[t!]
\centering
\vspace{-15mm}
\hspace{-5mm}\includegraphics[scale=0.45]{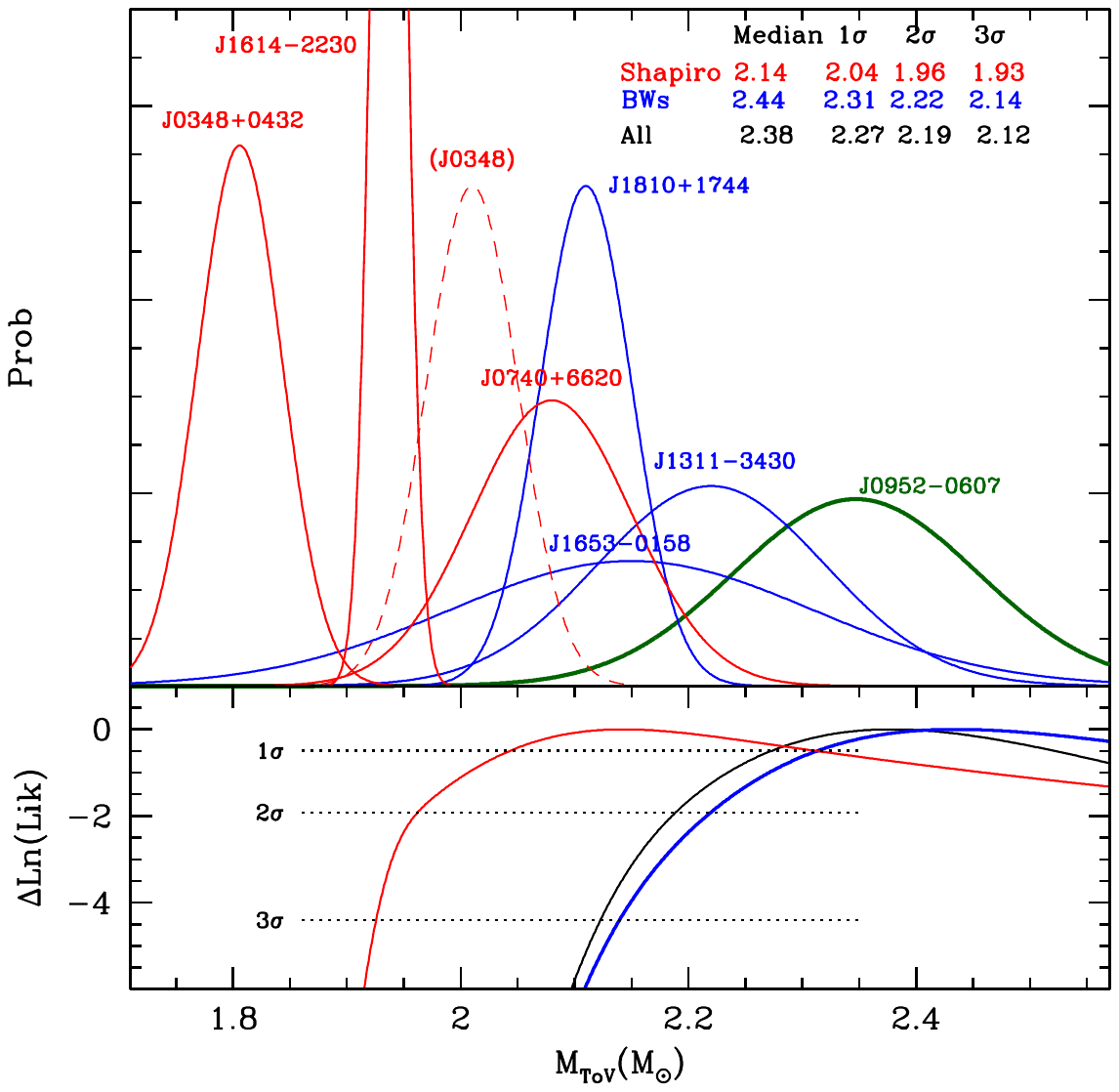}
\vspace{-29mm}
\caption{Mass estimates for heavy NSs. Three Shapiro-delay-measured MS/WD binaries are shown in red. Three black widows are in blue, with this paper's J0952 measurement in  green. The bottom panel shows the normalized Ln(Likelihood) for various combinations of these measurements, assuming a flat distribution of masses from 1.8\,$M_\odot$ up to some $M_{\rm ToV}$. The inset legend gives the median estimator for $M_{\rm ToV}$ as well as the $1\sigma$, $2\sigma$, and $3\sigma$ lower bounds on its value, for various millisecond pulsar sets.}
\label{fig:masses}
\end{figure}

\section{Mass Implications and Conclusions}

Here we revisit the minimum required $M_{\rm ToV}$ by plotting the Gaussian probability distribution functions with ${M_i,\,\sigma_i}$ for observed heavy NSs in Figure \ref{fig:masses}. Three millisecond-pulsar--WD binaries with Shapiro-delay measurements of $>1.8\,M_\odot$ are shown with the red solid Gaussians. These are J0348+0432 at $1.806\pm0.037\,M_\odot$ \citep{2025ApJ...983L..20S}, J1614$-$2230 at $1.937\pm0.014\,M_\odot$ \citep{2023ApJ...951L...9A}, and J0740+6620 at $2.08\pm0.07\,M_\odot$ \citep{2021ApJ...915L..12F}. The dashed Gaussian shows the $2.01\pm0.04\,M_\odot$ estimate for J0348+0432 \citep{antoniadis2013massive}. In blue we show three black widows measured by \citet{2023ApJ...942....6K}: J1810+1744 at $2.11\pm0.04\,M_\odot$, J1653$-$0158 at $2.15\pm0.16\,M_\odot$, and J1311$-$3430 at $2.22\pm0.10\,M_\odot$. Green indicates the J0952 measurement in this paper.

As in R22, we estimate the minimum value of $M_{\rm ToV}$ required by these observations by assuming that the NS mass distribution is flat $M^0$ above $M_1=1.8\,M_\odot$ to some cutoff value (for this $>1.8\,M_\odot$ sample, results are insensitive to the underlying distribution; values change by $<0.01\,M_\odot$ for $M^{-1}$). The log(likelihood) for $n$ Gaussian measurements is
\begin{equation}
\notag
\begin{split}
{\rm ln}\mathcal{L}= -&n\,{\rm ln}(M_{\rm max}-M_1) \\ 
+&\Sigma_{i=1}^n{\rm ln}\left [{\rm erf} \left( {\frac{M_i-M_1}{2^{1/2}\sigma_i}} \right ) -{\rm erf} \left( {\frac{M_i-M_{\rm max}}{2^{1/2} \sigma_i}}\right )
\right ].
\end{split}
\end{equation}  
These curves are plotted in the lower panel of Figure \ref{fig:masses}. We list the medians and lower bounds at various confidence levels on $M_{\rm max}$, from the likelihood ratio, for the different sample sets.

These estimates ignore the effect of rotational support, which can be important for high-spin NSs with very precise mass measurements; since J0952 is the second fastest rotator known, one expects that its central density will match that of a nonrotating star with somewhat lower mass. \cite{2022ApJ...934..139K} provide convenient equations relating rotating $M$ and nonrotating $M_*$ masses in terms of the compactness $C_*=M_*/R_*$. One can forward model for an assumed EoS and apply corrections for $C_*(M_*)$, or use individual measured $R$ values. While analysis using all EoS constraints is beyond the scope of this paper and while we lack $R$s for our pulsar sample, we can illustrate the corrections by assuming a reasonable $R_*$ for these massive NSs. Conservatively adopting $R_*=12$\,km, these equations indicate that the equivalent nonrotating J0952 mass is lower by $0.026\,M_\odot$ ($0.25\sigma$). For smaller radii the correction is smaller. The rotation correction is actually more important for J1810 and J1614, given their higher mass precision, but even these are small ($<0.6\sigma$). Still, as higher mass precision is achieved, such corrections should be applied.

The improved J0952 accuracy drives an ``all pulsar'' $1\sigma$ lower limit $M_{\rm ToV}=2.27\,M_\odot$, increased by $0.08\,M_\odot$ over that of R22; the $2.12\,M_\odot$ $3\sigma$ limit increases by $0.03\,M_\odot$. For the Shapiro-delay-constrained pure radio results one has a 1$\sigma$ lower bound just above $2\,M_\odot$, while the $3\sigma$ limit of $1.93\,M_\odot$ is now very well constrained by the highly precise J1614 mass measurement, as evidenced by the sharp lower cutoff in the likelihood function. A rotation correction, such as for $R_*=12$\,km, decreases these lower bounds, and those in Fig.\,\ref{fig:masses}, by an average of $0.015\,M_\odot$. Note that the ``Shapiro'' objects all have WD companions whose progenitors were well evolved at the end of mass transfer, driving high ${\dot M}$ and limiting the NS's mass acceptance. The black widows with very low-mass companions and irradiation-driven evolution can in principle reach higher masses. This drives our quest to measure their masses as accurately as possible and with minimal systematic/model-induced bias. 

Of course, the Shapiro measurements are essentially free from systematics, so the  cautious theorist should adopt these bounds for EoS constraints. Measurements involving optical modeling are more subject to error. J0348 is an object lesson; its earlier $2.01\,M_\odot$ mass measurement relied on optical spectroscopic estimates of the WD companion's mass. While statistically precise, these were evidently biased, and now the revised radio-only J0348 measurement is $\sim 5\sigma$ lower and contributes essentially nothing to the $M_{\rm ToV}$ bound. Our J0952 measurement might be subject to similar issues. With our new photometry we feel that the light-curve fits are quite robust; barring a major revision of our understanding of companion heating, the inclination estimates seem unlikely to change. However, the significant scatter and outliers in the RV measurements of Fig.\,\ref{fig:RVfits} make this component less secure. Indeed, the need to inspect the cross-correlation functions to isolate the RV peak inevitably introduces some subjectivity in the measurements, insufficiently suppressed by the synthesis of the model template spectra. This synthesis has, on the other hand, retired the systematic uncertainty associated with the $K_{\rm cor}$ modeling. 

However, if black widows {\it can} achieve higher masses, it remains imperative to pin down their parameters. Unfortunately, with low companion masses and with the evaporation driving a significant ionized wind, radio Shapiro techniques cannot be applied. Of the black widows discussed here, J0952 is the simplest, but unfortunately the faintest  and observationally most challenging, target. J1810 is bright and well measured but has evidence for a companion hot spot, which complicates modeling. J1311 is also bright, but with extremely strong pulsar heating suffers frequent and bright companion flares, making measurement of the true quiescence light curve challenging. J1653, the Galactic black widow with the shortest orbital period, has a bright and variable veiling flux component covering binary minimum brightness, challenging modeling and limiting precision of the $i$ measurements. Also, J1311 and J1653 are members of the Tidarren black widow subclass, with H-free companion surfaces, making spectral modeling and CoM RV determination difficult. Similarly, the Tidarren PSR J2322$-$2650 shows a beautiful RV curve in new {\it JWST} data \citep{2025arXiv250904558Z}, but has a very low-temperature companion dominated by molecular carbon, so modeling has not yet been able to yield a precise mass. J0952, with its normal H-dominated surface, weak heating, modest Roche lobe fill factor, and long orbital period, remains the most securely modeled black widow. While we certainly expect more black-widow discoveries, especially from follow-up observations of {\it Fermi} sources, these will be increasingly distant and (likely) optically faint. Thus, the safest path to progress lies with better spectroscopy of J0952, as this now limits the (statistical) mass precision. This would require significant additional investment of large-aperture telescope time. More efficient spectrographs, perhaps including integral field units to allow improved relative spectrophotometry versus stable comparison stars, can help. But J0952 remains at the limit of the possible, and for higher precision measurements we may need to look to the next generation of large optical telescopes.

\bigskip

This paper's data were obtained at the W.~M. Keck Observatory, which is operated as a scientific partnership among the California Institute of Technology, the University of California, and NASA; it was made possible by the generous financial support of the W.~M. Keck Foundation. We are grateful for the excellent assistance of the observatory staff, as well as the Keck Time Allocation Committee and Keck scheduler for accommodating our requests for specific half-nights.
%M.B. and R.W.R. were supported in part by NASA grants 80NSSC17K0024 and 80NSSC17K0502.                        
A.V.F.’s team at U.C. Berkeley received support from the Christopher R. Redlich 
Fund, Gary and Cynthia Bengier, Clark and Sharon Winslow, Alan 
Eustace and Kathy Kwan (W.Z. is a Bengier-Winslow-Eustace Specialist in       
Astronomy), William Draper, Timothy and Melissa Draper, Briggs                
and Kathleen Wood, Sanford Robertson (T.G.B. is a                
Draper-Wood-Robertson Specialist in Astronomy), 
and many other donors.

\bibliographystyle{aasjournal}
\bibliography{J0952_main}

@article{alsing2018evidence,
  title={Evidence for a maximum mass cut-off in the neutron star mass distribution and constraints on the equation of state},
  author={Alsing, Justin and Silva, Hector O and Berti, Emanuele},
  journal={\mnras},
  volume={478},
  number={1},
  pages={1377--1391},
  year={2018},
  publisher={Oxford University Press}
}

@article{oke1995keck,
  title={The Keck low-resolution imaging spectrometer},
  author={Oke, JB and Cohen, JG and Carr, M and Cromer, J and Dingizian, A and Harris, FH and Labrecque, S and Lucinio, R and Schaal, W and Epps, H and others},
  journal={Publications of the Astronomical Society of the Pacific},
  volume={107},
  number={710},
  pages={375},
  year={1995},
  publisher={IOP Publishing}
}

@article{antoniadis2013massive,
  title={A massive pulsar in a compact relativistic binary},
  author={Antoniadis, John and Freire, Paulo CC and Wex, Norbert and Tauris, Thomas M and Lynch, Ryan S and van Kerkwijk, Marten H and Kramer, Michael and G, Cees and Dhillon, Vik S and Driebe, Thomas and others},
  journal={Science},
  volume={340},
  number={6131},
  year={2013},
  publisher={American Association for the Advancement of Science}
}

@article{finkbeiner2016hypercalibration,
  title={Hypercalibration: a Pan-STARRS1-based recalibration of the Sloan Digital Sky Survey photometry},
  author={Finkbeiner, Douglas P and Schlafly, Edward F and Schlegel, David J and Padmanabhan, Nikhil and Juri{\'c}, Mario and Burgett, William S and Chambers, Kenneth C and Denneau, Larry and Draper, Peter W and Flewelling, Heather and others},
  journal={\apj},
  volume={822},
  number={2},
  pages={66},
  year={2016},
  publisher={IOP Publishing}
}

@article{breton2012koi,
  title={KOI 1224: a fourth bloated hot white dwarf companion found with Kepler},
  author={Breton, Rene P and Rappaport, Saul A and van Kerkwijk, Marten H and Carter, Josh A},
  journal={\apj},
  volume={748},
  number={2},
  pages={115},
  year={2012},
  publisher={IOP Publishing}
}

@article{Kandel_2020,
	doi = {10.3847/1538-4357/ab7b62},
	year = 2020,
	month = {apr},
	publisher = {American Astronomical Society},
	volume = {892},
	number = {2},
	pages = {101},
	author = {D. Kandel and Roger W. Romani},
	title = {Atmospheric Circulation on Black Widow Companions},
	journal = {\apj}
}

@ARTICLE{2017ApJ...845...42S,
   author = {{Sanchez}, N. and {Romani}, R.~W.},
    title = "{B-ducted Heating of Black Widow Companions}",
  journal = {\apj},
archivePrefix = "arXiv",
   eprint = {1706.05467},
 primaryClass = "astro-ph.HE",
     year = 2017,
    month = aug,
   volume = 845,
      eid = {42},
    pages = {42},
      doi = {10.3847/1538-4357/aa7a02},
   adsurl = {http://adsabs.harvard.edu/abs/2017ApJ...845...42S}
}

@article{romani2011orbit,
  title={The orbit and companion of probable $\gamma$-ray pulsar J2339--0533},
  author={Romani, Roger W and Shaw, Michael S},
  journal={\apjl},
  volume={743},
  number={2},
  pages={L26},
  year={2011},
  publisher={IOP Publishing}
}

@ARTICLE{2017ApJ...846L..20B,
       author = {{Bassa}, C.~G. and {Pleunis}, Z. and {Hessels}, J.~W.~T. and {Ferrara}, E.~C. and {Breton}, R.~P. and {Gusinskaia}, N.~V. and {Kondratiev}, V.~I. and {Sanidas}, S. and {Nieder}, L. and {Clark}, C.~J. and {Li}, T. and {van Amesfoort}, A.~S. and {Burnett}, T.~H. and {Camilo}, F. and {Michelson}, P.~F. and {Ransom}, S.~M. and {Ray}, P.~S. and {Wood}, K.},
        title = "{LOFAR Discovery of the Fastest-spinning Millisecond Pulsar in the Galactic Field}",
      journal = {\apjl},
         year = 2017,
        month = sep,
       volume = {846},
       number = {2},
          eid = {L20},
        pages = {L20},
          doi = {10.3847/2041-8213/aa8400},
archivePrefix = {arXiv},
       eprint = {1709.01453},
 primaryClass = {astro-ph.HE},
       adsurl = {https://ui.adsabs.harvard.edu/abs/2017ApJ...846L..20B}
}

@ARTICLE{2019ApJ...883...42N,
       author = {{Nieder}, L. and {Clark}, C.~J. and {Bassa}, C.~G. and {Wu}, J. and {Singh}, A. and {Donner}, J.~Y. and {Allen}, B. and {Breton}, R.~P. and {Dhillon}, V.~S. and {Eggenstein}, H. -B. and {Hessels}, J.~W.~T. and {Kennedy}, M.~R. and {Kerr}, M. and {Littlefair}, S. and {Marsh}, T.~R. and {Mata S{\'a}nchez}, D. and {Papa}, M.~A. and {Ray}, P.~S. and {Steltner}, B. and {Verbiest}, J.~P.~W.},
        title = "{Detection and Timing of Gamma-Ray Pulsations from the 707 Hz Pulsar J0952-0607}",
      journal = {\apj},
         year = 2019,
        month = sep,
       volume = {883},
       number = {1},
          eid = {42},
        pages = {42},
          doi = {10.3847/1538-4357/ab357e},
archivePrefix = {arXiv},
       eprint = {1905.11352},
 primaryClass = {astro-ph.HE},
       adsurl = {https://ui.adsabs.harvard.edu/abs/2019ApJ...883...42N}
}

@ARTICLE{2021ApJ...908L..46R,
       author = {{Romani}, Roger W. and {Kandel}, D. and {Filippenko}, Alexei V. and {Brink}, Thomas G. and {Zheng}, WeiKang},
        title = "{PSR J1810+1744: Companion Darkening and a Precise High Neutron Star Mass}",
      journal = {\apjl},
         year = 2021,
        month = feb,
       volume = {908},
       number = {2},
          eid = {L46},
        pages = {L46},
          doi = {10.3847/2041-8213/abe2b4},
archivePrefix = {arXiv},
       eprint = {2101.09822},
 primaryClass = {astro-ph.HE},
       adsurl = {https://ui.adsabs.harvard.edu/abs/2021ApJ...908L..46R}
}

@article{feroz2009multinest,
  title={MultiNest: an efficient and robust Bayesian inference tool for cosmology and particle physics},
  author={Feroz, Farhan and Hobson, MP and Bridges, Michael},
  journal={Monthly Notices of the Royal Astronomical Society},
  volume={398},
  number={4},
  pages={1601--1614},
  year={2009},
  publisher={Blackwell Publishing Ltd Oxford, UK}
}

@ARTICLE{2023ApJ...942....6K,
       author = {{Kandel}, D. and {Romani}, Roger W.},
        title = "{An Optical Study of the Black Widow Population}",
      journal = {\apj},
         year = 2023,
        month = jan,
       volume = {942},
       number = {1},
          eid = {6},
        pages = {6},
          doi = {10.3847/1538-4357/aca524},
archivePrefix = {arXiv},
       eprint = {2211.16990},
 primaryClass = {astro-ph.HE},
       adsurl = {https://ui.adsabs.harvard.edu/abs/2023ApJ...942....6K}
}

@article{buchner2016pymultinest,
  title={PyMultiNest: Python interface for MultiNest},
  author={Buchner, Johannes},
  journal={Astrophysics Source Code Library},
  pages={ascl--1606},
  year={2016}
}

@ARTICLE{1970SvA....13..562S,
       author = {{Shklovskii}, I.~S.},
        title = "{Possible Causes of the Secular Increase in Pulsar Periods.}",
      journal = {\sovast},
         year = 1970,
        month = feb,
       volume = {13},
        pages = {562},
       adsurl = {https://ui.adsabs.harvard.edu/abs/1970SvA....13..562S}
}

@ARTICLE{1998PASP..110..934K,
       author = {{Kurtz}, Michael J. and {Mink}, Douglas J.},
        title = "{RVSAO 2.0: Digital Redshifts and Radial Velocities}",
      journal = {\pasp},
         year = 1998,
        month = aug,
       volume = {110},
       number = {750},
        pages = {934-977},
          doi = {10.1086/316207},
archivePrefix = {arXiv},
       eprint = {astro-ph/9803252},
 primaryClass = {astro-ph},
       adsurl = {https://ui.adsabs.harvard.edu/abs/1998PASP..110..934K}
}

@ARTICLE{2021ApJ...915L..12F,
       author = {{Fonseca}, E. and {Cromartie}, H.~T. and {Pennucci}, T.~T. and {Ray}, P.~S. and {Kirichenko}, A. Yu. and {Ransom}, S.~M. and {Demorest}, P.~B. and {Stairs}, I.~H. and {Arzoumanian}, Z. and {Guillemot}, L. and {Parthasarathy}, A. and {Kerr}, M. and {Cognard}, I. and {Baker}, P.~T. and {Blumer}, H. and {Brook}, P.~R. and {DeCesar}, M. and {Dolch}, T. and {Dong}, F.~A. and {Ferrara}, E.~C. and {Fiore}, W. and {Garver-Daniels}, N. and {Good}, D.~C. and {Jennings}, R. and {Jones}, M.~L. and {Kaspi}, V.~M. and {Lam}, M.~T. and {Lorimer}, D.~R. and {Luo}, J. and {McEwen}, A. and {McKee}, J.~W. and {McLaughlin}, M.~A. and {McMann}, N. and {Meyers}, B.~W. and {Naidu}, A. and {Ng}, C. and {Nice}, D.~J. and {Pol}, N. and {Radovan}, H.~A. and {Shapiro-Albert}, B. and {Tan}, C.~M. and {Tendulkar}, S.~P. and {Swiggum}, J.~K. and {Wahl}, H.~M. and {Zhu}, W.~W.},
        title = "{Refined Mass and Geometric Measurements of the High-mass PSR J0740+6620}",
      journal = {\apjl},
         year = 2021,
        month = jul,
       volume = {915},
       number = {1},
          eid = {L12},
        pages = {L12},
          doi = {10.3847/2041-8213/ac03b8},
archivePrefix = {arXiv},
       eprint = {2104.00880},
 primaryClass = {astro-ph.HE},
       adsurl = {https://ui.adsabs.harvard.edu/abs/2021ApJ...915L..12F}
}

@ARTICLE{2022ApJ...934L..17R,
       author = {{Romani}, Roger W. and {Kandel}, D. and {Filippenko}, Alexei V. and {Brink}, Thomas G. and {Zheng}, WeiKang},
        title = "{PSR J0952-0607: The Fastest and Heaviest Known Galactic Neutron Star}",
      journal = {\apjl},
         year = 2022,
        month = aug,
       volume = {934},
       number = {2},
          eid = {L17},
        pages = {L17},
          doi = {10.3847/2041-8213/ac8007},
archivePrefix = {arXiv},
       eprint = {2207.05124},
 primaryClass = {astro-ph.HE},
       adsurl = {https://ui.adsabs.harvard.edu/abs/2022ApJ...934L..17R}
}

@ARTICLE{2013MSAIS..24..128A,
       author = {{Allard}, F. and {Homeier}, D. and {Freytag}, B. and {Schaffenberger}, W. and {Rajpurohit}, A.~S.},
        title = "{Progress in modeling very low mass stars, brown dwarfs, and planetary mass objects.}",
      journal = {Memorie della Societa Astronomica Italiana Supplementi},
         year = 2013,
        month = jan,
       volume = {24},
        pages = {128},
          doi = {10.48550/arXiv.1302.6559},
archivePrefix = {arXiv},
       eprint = {1302.6559},
 primaryClass = {astro-ph.SR},
       adsurl = {https://ui.adsabs.harvard.edu/abs/2013MSAIS..24..128A}
}

@ARTICLE{2025ApJ...983L..20S,
       author = {{Saffer}, Alexander and {Fonseca}, Emmanuel and {Ransom}, Scott and {Stairs}, Ingrid and {Lynch}, Ryan and {Good}, Deborah and {Masui}, Kiyoshi W. and {McKee}, James W. and {Meyers}, Bradley W. and {Patil}, Swarali Shivraj and {Tan}, Chia Min},
        title = "{A Lower Mass Estimate for PSR J0348+0432 Based on CHIME/Pulsar Precision Timing}",
      journal = {\apjl},
         year = 2025,
        month = apr,
       volume = {983},
       number = {1},
          eid = {L20},
        pages = {L20},
          doi = {10.3847/2041-8213/adc25e},
archivePrefix = {arXiv},
       eprint = {2412.02850},
 primaryClass = {astro-ph.HE},
       adsurl = {https://ui.adsabs.harvard.edu/abs/2025ApJ...983L..20S}
}

@ARTICLE{2023ApJ...951L...9A,
       author = {{Agazie}, Gabriella and {Alam}, Md Faisal and {Anumarlapudi}, Akash and {Archibald}, Anne M. and {Arzoumanian}, Zaven and {Baker}, Paul T. and {Blecha}, Laura and {Bonidie}, Victoria and {Brazier}, Adam and {Brook}, Paul R. and {Burke-Spolaor}, Sarah and {B{\'e}csy}, Bence and {Chapman}, Christopher and {Charisi}, Maria and {Chatterjee}, Shami and {Cohen}, Tyler and {Cordes}, James M. and {Cornish}, Neil J. and {Crawford}, Fronefield and {Cromartie}, H. Thankful and {Crowter}, Kathryn and {Decesar}, Megan E. and {Demorest}, Paul B. and {Dolch}, Timothy and {Drachler}, Brendan and {Ferrara}, Elizabeth C. and {Fiore}, William and {Fonseca}, Emmanuel and {Freedman}, Gabriel E. and {Garver-Daniels}, Nate and {Gentile}, Peter A. and {Glaser}, Joseph and {Good}, Deborah C. and {G{\"u}ltekin}, Kayhan and {Hazboun}, Jeffrey S. and {Jennings}, Ross J. and {Jessup}, Cody and {Johnson}, Aaron D. and {Jones}, Megan L. and {Kaiser}, Andrew R. and {Kaplan}, David L. and {Kelley}, Luke Zoltan and {Kerr}, Matthew and {Key}, Joey S. and {Kuske}, Anastasia and {Laal}, Nima and {Lam}, Michael T. and {Lamb}, William G. and {Lazio}, T. Joseph W. and {Lewandowska}, Natalia and {Lin}, Ye and {Liu}, Tingting and {Lorimer}, Duncan R. and {Luo}, Jing and {Lynch}, Ryan S. and {Ma}, Chung-Pei and {Madison}, Dustin R. and {Maraccini}, Kaleb and {McEwen}, Alexander and {McKee}, James W. and {McLaughlin}, Maura A. and {McMann}, Natasha and {Meyers}, Bradley W. and {Mingarelli}, Chiara M.~F. and {Mitridate}, Andrea and {Ng}, Cherry and {Nice}, David J. and {Ocker}, Stella Koch and {Olum}, Ken D. and {Panciu}, Elisa and {Pennucci}, Timothy T. and {Perera}, Benetge B.~P. and {Pol}, Nihan S. and {Radovan}, Henri A. and {Ransom}, Scott M. and {Ray}, Paul S. and {Romano}, Joseph D. and {Salo}, Laura and {Sardesai}, Shashwat C. and {Schmiedekamp}, Carl and {Schmiedekamp}, Ann and {Schmitz}, Kai and {Shapiro-Albert}, Brent J. and {Siemens}, Xavier and {Simon}, Joseph and {Siwek}, Magdalena S. and {Stairs}, Ingrid H. and {Stinebring}, Daniel R. and {Stovall}, Kevin and {Susobhanan}, Abhimanyu and {Swiggum}, Joseph K. and {Taylor}, Stephen R. and {Turner}, Jacob E. and {Unal}, Caner and {Vallisneri}, Michele and {Vigeland}, Sarah J. and {Wahl}, Haley M. and {Wang}, Qiaohong and {Witt}, Caitlin A. and {Young}, Olivia and {Nanograv Collaboration}},
        title = "{The NANOGrav 15 yr Data Set: Observations and Timing of 68 Millisecond Pulsars}",
      journal = {\apjl},
         year = 2023,
        month = jul,
       volume = {951},
       number = {1},
          eid = {L9},
        pages = {L9},
          doi = {10.3847/2041-8213/acda9a},
archivePrefix = {arXiv},
       eprint = {2306.16217},
 primaryClass = {astro-ph.HE},
       adsurl = {https://ui.adsabs.harvard.edu/abs/2023ApJ...951L...9A}
}

@ARTICLE{2025arXiv250904558Z,
       author = {{Zhang}, Michael and {Beleznay}, Maya and {Brandt}, Timothy D. and {Romani}, Roger W. and {Gao}, Peter and {Beltz}, Hayley and {Bailes}, Matthew and {Nixon}, Matthew C. and {Bean}, Jacob L. and {Komacek}, Thaddeus D. and {Coy}, Brandon P. and {Fu}, Guangwei and {Luque}, Rafael and {Reardon}, Daniel J. and {Carli}, Emma and {Shannon}, Ryan M. and {Fortney}, Jonathan J. and {Piette}, Anjali A.~A. and {Miller}, M. Coleman and {Desert}, Jean-Michel},
        title = "{A carbon-rich atmosphere on a windy pulsar planet}",
      journal = {arXiv e-prints},
         year = 2025,
        month = sep,
          eid = {arXiv:2509.04558},
        pages = {arXiv:2509.04558},
          doi = {10.48550/arXiv.2509.04558},
archivePrefix = {arXiv},
       eprint = {2509.04558},
 primaryClass = {astro-ph.EP},
       adsurl = {https://ui.adsabs.harvard.edu/abs/2025arXiv250904558Z}
}

@ARTICLE{2022ApJ...934..139K,
       author = {{Konstantinou}, Andreas and {Morsink}, Sharon M.},
        title = "{Universal Relations for the Increase in the Mass and Radius of a Rotating Neutron Star}",
      journal = {\apj},
         year = 2022,
        month = aug,
       volume = {934},
       number = {2},
          eid = {139},
        pages = {139},
          doi = {10.3847/1538-4357/ac7b86},
archivePrefix = {arXiv},
       eprint = {2206.12515},
 primaryClass = {astro-ph.HE},
       adsurl = {https://ui.adsabs.harvard.edu/abs/2022ApJ...934..139K}
}
\end{document}